# Magnetic anisotropy of ferromagnetic metals in low-symmetry systems

Yoshishige Suzuki[1,2,3*] and Shinji Miwa[4,2]

[1]*Graduate School of Engineering Science, Osaka University, Toyonaka, Osaka 560-8531, Japan*

[2]*Center for Spintronics Research Network (CSRN), Osaka University, Toyonaka, Osaka 560-8531, Japan*

[3]*Materials research by information Integration Initiative (MI²I) project, National Institute for Materials Science (NIMS), Tsukuba, Ibaraki 305-0047, Japan*

[4]*The Institute for Solid State Physics, The University of Tokyo, Kashiwa, Chiba 277-8581, Japan*

[*]E-mail: suzuki-y@mp.es.osaka-u.ac.jp

We have constructed an analytic formula to treat the perpendicular magnetic anisotropy energy in ferromagnetic metals with low symmetry, such as $C_{4V}$ and $C_{3V}$. We find that the anisotropy energy is proportional to a part of the expectation values of the orbital angular momentum and magnetic dipole operator. Although the result is similar to the model proposed by Laan [J. Phys.: Condens. Matter **10**, 3239 (1998)], we have derived a concrete expression for the spin-flip virtual excitation process term, which can be dominant in atoms with small magnetic moments and/or small exchange splitting. Pt monatomic layer with proximity-induced spin polarization grown on Fe is an example of this. Other multilayer systems such as Co/Pd and Co/Ni and bilayer systems such as Fe(CoB)/MgO can be discussed similarly. Moreover, the relation between perpendicular magnetic anisotropy energy and measurable physical parameters is discussed based on X-ray magnetic circular dichroism spectroscopy.

The theory for perpendicular magnetic anisotropy (PMA) in ferromagnetic ultrathin metals[1-3] has been intensively developed. The physical origin of the PMA in ultrathin metals is correlated with its low symmetry, and is essentially the same as that for uniaxial anisotropy in bulk materials.[4] It is still important to understand the physical mechanism of the PMA[5,6] because it is necessary to employ perpendicularly magnetized films for high-density magnetic recording disk and magnetic random access memory (MRAM) devices. Moreover, the PMA can be controlled by an external electric field,[6-8] which is called the voltage-controlled magnetic anisotropy (VCMA) effect. The VCMA-induced voltage-driven torque[9] can replace the current-driven spin-transfer torque in an MRAM application because of its ultralow power consumption. Because an electric field is shielded at the surface/interface of metals, VCMA is a surface/interface effect and requires deeper understanding of the physics of PMA.[10,20]

One mechanism to explain PMA is the magnetization direction dependence of its orbital magnetic moment because the spin-orbit interaction energy is proportional to the size of the orbital angular momentum. This relation is known as the Bruno model.[1] When we treat the second order perturbation in the spin-orbit interaction energy, the Bruno relation can be derived from the spin-conserved (non-flip) virtual excitation processes. Besides this, it has been pointed out that the PMA energy is also correlated with the intra-atomic magnetic dipole operator through spin-flip virtual excitation processes.[2,3] In this paper, we have treated the second order perturbation in the spin-orbit interaction energy in a different manner. Although the result is similar to the previously proposed model,[2] we have derived a concrete expression of the spin-flip terms, which can be dominant in atoms with small magnetic moments and/or small exchange splitting. Pt monatomic layer with proximity induced spin polarization grown on Fe is an example.[10] Other multilayer PMA systems such as Co/Pd[5] and Co/Ni[11-13] and bilayer PMA systems such as Fe(CoB)/MgO[14-20] can be discussed similarly. Lately, the importance of spin-flip virtual excitation terms has been discussed not only for strong spin-orbit interaction systems such as Fe/Pt[10] but also for weak spin-orbit interaction systems such as Fe/MgO[19,21] and Co/Ni.[13] Toward the end of this paper, the relation between PMA energy and the measurable physical parameters is discussed based on X-ray magnetic circular dichroism spectroscopy.

In this paper, we treat only ferromagnetic metals made of a single type of atoms for simplicity. The system has uniaxial symmetry like $C_{4v}$ or $C_{3v}$. For example, Pt monatomic layer in contact with bulk Fe and a pure ferromagnetic metal with distorted lattice can be treated with the model. The model only includes $d$-electrons by considering the atomic orbital at each atom $|j,\mu,\sigma\rangle$ as the basis. Here, $j$ is the atomic position, $\sigma = \pm 1/2$ is the spin quantum number, and $\mu$ is the quantum number for the $z$-component of the angular momentum operator, i.e., $L_z|j,\mu,\sigma\rangle = \mu\hbar|j,\mu,\sigma\rangle$. The phase of the spherical harmonics is the same as that employed in Ref. 22. Except for the basis defined above, we follow the expressions presented in Ref. 1. The Bloch state of the eigenenergy $\varepsilon_{n,\sigma}(\mathbf{k})$ is $|\mathbf{k},n,\sigma\rangle$. The eigenstate is constructed from the atomic orbitals as

$$|\mathbf{k},n,\sigma\rangle = \frac{1}{\sqrt{N}} \sum_{\mu,j} e^{i\mathbf{k}\cdot\mathbf{x}_j} a_{n,\mu,\sigma}(\mathbf{k}) |j,\mu,\sigma\rangle. \tag{1}$$

Here, $N$ is the number of atoms in the crystal and $a_{n,\mu,\sigma}(\mathbf{k})$ is a coefficient of the development. We express the spin-orbit interaction Hamiltonian ($\hat{\mathcal{H}}_{\text{SOI}}$) using the basis defined above. The spin-orbit interaction Hamiltonian is

$$\begin{aligned}\mathcal{H}_{\text{SOI}} &= \frac{\xi}{\hbar^2} \sum_{\substack{n_2,n_1,\sigma_2,\sigma_1 \\ \mathbf{k}_1,\mathbf{k}_2}} \langle \mathbf{k}_2,n_2,\sigma_2|\mathbf{L}\cdot\mathbf{S}|\mathbf{k}_1,n_1,\sigma_1\rangle c^\dagger_{n_2,\sigma_2}(\mathbf{k}_2) c_{n_1,\sigma_1}(\mathbf{k}_1) \\ &= \frac{\xi}{\hbar^2} \sum_{\mu_2,\mu_1,\sigma_2,\sigma_1} \langle \mu_2,\sigma_2|\mathbf{L}\cdot\mathbf{S}|\mu_1,\sigma_1\rangle \sum_{\substack{n_2,n_1 \\ \mathbf{k}}} a^*_{n_2,\mu_2,\sigma_2}(\mathbf{k}) a_{n_1,\mu_1,\sigma_1}(\mathbf{k}) c^\dagger_{n_2,\sigma_2}(\mathbf{k}) c_{n_1,\sigma_1}(\mathbf{k}),\end{aligned} \tag{2}$$

where $\xi$ is the spin-orbit interaction coefficient. The matrix elements are evaluated at the same atomic site by integrating spherical coordinate angles $\theta$ and $\phi$. The radial part of the integration is common and included in the normalization factor. The lowest order correction to the energy and the ground state vector are

$$\begin{cases} \delta E = \sum_{\text{exc}} \dfrac{|\langle \text{gr}|\mathcal{H}_{\text{SOI}}|\text{exc}\rangle|^2}{E_{\text{gr}} - E_{\text{exc}}} \\ \delta|\text{gr}\rangle = \sum_{\text{exc}} \dfrac{|\text{exc}\rangle\langle \text{exc}|\mathcal{H}_{\text{SOI}}|\text{gr}\rangle}{E_{\text{gr}} - E_{\text{exc}}} \end{cases}. \tag{3}$$

Using the orbital expression of the spin-orbit expression shown in Eq. (2), the energy correction is expressed as[2]

$$\delta E = \frac{\xi^2}{\hbar^4} \sum_{\theta,\sigma_2,\sigma_1} \langle \mu_1',\sigma_1|\mathbf{L}\cdot\mathbf{S}|\mu_2',\sigma_2\rangle\langle \mu_2,\sigma_2|\mathbf{L}\cdot\mathbf{S}|\mu_1,\sigma_1\rangle A(\theta,\sigma_2,\sigma_1). \tag{4}$$

Here, $\theta$ represents the set of indices $\theta = (\mu_1',\mu_2',\mu_2,\mu_1)$. The generalized density of states $n_{\mu',\mu,\sigma}(\mathbf{k},\varepsilon)$, generalized joint-density of states $J(\theta,\sigma_2,\sigma_1,\varepsilon)$, and the factor $A(\theta,\sigma_2,\sigma_1)$ are defined below.

$$\begin{aligned} n_{\mu',\mu,\sigma}(\mathbf{k},\varepsilon) &\equiv \sum_n a^*_{n,\mu',\sigma}(\mathbf{k})a_{n,\mu,\sigma}(\mathbf{k})\delta(\varepsilon - \varepsilon_{n,\sigma}(\mathbf{k})) \\ J(\theta,\sigma_2,\sigma_1,\varepsilon) &\equiv \sum_{\mathbf{k}} \int^{\varepsilon_F} d\varepsilon_1 n_{\mu_2,\mu_2',\sigma_2}(\mathbf{k},\varepsilon_1+\varepsilon) n_{\mu_1',\mu_1,\sigma_1}(\mathbf{k},\varepsilon_1)H(\varepsilon_1+\varepsilon-\varepsilon_F) \\ A(\theta,\sigma_2,\sigma_1) &\equiv -\int_{+0} d\varepsilon \frac{J(\theta,\sigma_2,\sigma_1,\varepsilon)}{\varepsilon} \end{aligned} \tag{5}$$

Here, $H(\varepsilon)$ is Heaviside step function. In Eq. (4), a virtually excited intermediate state may have parallel or anti-parallel spin with regard to the initial state in the ground state. Hereafter, we call those contributions spin-conserved and spin-flip virtual excitation processes.

Next, we derive the relation between spin-orbit interaction energy and orbital angular momentum $\langle \mathbf{L} \rangle$. The expectation value of the orbital angular momentum for the perturbed ground state can be obtained as follows:

$$\begin{cases} \langle \mathbf{L} \rangle = -\sum_{\text{exc}} \frac{\langle \text{gr}|\mathcal{H}_{\text{SOI}}|\text{exc}\rangle\langle \text{exc}|\mathbf{L}|\text{gr}\rangle}{E_{\text{exc}} - E_{\text{gr}}} + \text{c.c.} \\ \mathbf{L} = \sum_{\sigma} \mathbf{L}_{\sigma\sigma} \\ \phantom{\mathbf{L}} = \sum_{\mu_2,\mu_1,\sigma} \langle \mu_2,\sigma|\mathbf{L}|\mu_1,\sigma\rangle \sum_{n_2,n_1,\mathbf{k}} a^*_{n_2,\mu_2,\sigma}(\mathbf{k}) a_{n_1,\mu_1,\sigma}(\mathbf{k}) c^\dagger_{n_2,\sigma}(\mathbf{k}) c_{n_1,\sigma}(\mathbf{k}) \end{cases} \quad (6)$$

It should be noted that $\mathbf{L}$ is diagonal for spin. Here, we consider the spin quantization axis ($z$-axis) parallel to the spin moment. We also assume that the induced orbital angular momentum is parallel to the spin angular momentum. Then, the following relation can be used (Eq. (4) in Ref. 1).

$$\langle \mu_2,\sigma_2|L_z|\mu_1,\sigma_1\rangle = \frac{1}{\sigma_1 \hbar} \delta_{\sigma_2,\sigma_1} \langle \mu_2,\sigma_2|\mathbf{L}\cdot\mathbf{S}|\mu_1,\sigma_1\rangle \quad (7)$$

By substituting Eq. (7) into Eq. (6), the contribution of the spin-conserved virtual excitation process to the spin-orbit interaction energy can be expressed using spin-resolved expectation values of the orbital angular momentum.[1,2] This relation is known as Bruno's relation.

$$\langle L_{z,\sigma\sigma}\rangle = -\frac{2}{\sigma\hbar\xi} \sum_{\text{exc}} \frac{\langle \text{gr}|\xi \mathbf{L}_{\sigma\sigma}\cdot\mathbf{S}|\text{exc}\rangle\langle \text{exc}|\mathcal{H}_{\text{SOI}}|\text{gr}\rangle}{E_{\text{exc}} - E_{\text{gr}}}$$
$$\therefore \delta E_{\text{spin-conserved}} = -\frac{\xi}{4\hbar}\left(\langle L_{z,\downarrow\downarrow}\rangle - \langle L_{z,\uparrow\uparrow}\rangle\right) \quad (8)$$

Next, we consider a relation between spin-orbit interaction energy and intra-atomic magnetic dipole operator ($\langle \mathbf{T}\rangle \equiv \langle \mathbf{S} - 3\mathbf{x}(\mathbf{x}\cdot\mathbf{S})\rangle = \langle \mathbf{Q}\cdot\mathbf{S}\rangle$). Here, $\mathbf{x}$ is the direction vector of unit length. $\mathbf{Q}$ is the dimensionless charge-quadrupole operator. For our basis, the matrix elements of $\mathbf{Q}$ are equal to that of the following operator constructed from the orbital angular momentum operator.[23,24]

$$Q_{\alpha\beta} = \delta_{\alpha\beta} - 3x_\alpha x_\beta = \frac{2}{7\hbar^2}\left(\frac{L_\alpha L_\beta + L_\beta L_\alpha}{2} - \frac{1}{3}L^2 \delta_{\alpha\beta}\right) \quad (9)$$

The expectation value of the magnetic dipole operator under spin-orbit interaction can be obtained for the first order approximation.

$$\langle \mathbf{T} \rangle \cong \left( \langle \mathrm{gr} | + \delta \langle \mathrm{gr} | \right) \mathbf{T} \left( | \mathrm{gr} \rangle + \delta | \mathrm{gr} \rangle \right)$$

$$\cong \langle \mathrm{gr} | \mathbf{T} | \mathrm{gr} \rangle + \left( \left( \delta \langle \mathrm{gr} | \right) \mathbf{T} | \mathrm{gr} \rangle + c.c. \right) \quad (10)$$

Here, $\langle \mathrm{gr} | \mathbf{T} | \mathrm{gr} \rangle$ may possess a finite value while $\langle \mathrm{gr} | \mathbf{L} | \mathrm{gr} \rangle$ is zero. In other words, when the ferromagnetic system possesses electric quadrupole, $\langle \mathbf{T} \rangle$ has a finite value even if the spin-orbit interaction is absent. From Eqs. (9) and (10), the following relation can be derived.

$$\langle \mathrm{gr} | T_z | \mathrm{gr} \rangle = -\frac{4}{7} \langle S_z \rangle + \frac{1}{7\hbar} \left( \langle L_{z,\uparrow\uparrow}^2 \rangle - \langle L_{z,\downarrow\downarrow}^2 \rangle \right) \quad (11)$$

Here, $\langle \mathrm{gr} | \mathbf{T} | \mathrm{gr} \rangle$ consists only of spin-conserved terms. It should be noted that $\langle \mathrm{gr} | \mathbf{T} | \mathrm{gr} \rangle$ may possess a finite value but it is not directly correlated with the PMA energy. To characterize the PMA energy, $\langle \delta \mathbf{T} \rangle \equiv \langle \mathrm{gr} | \mathbf{T} \left( \delta | \mathrm{gr} \rangle \right) + \left( \delta \langle \mathrm{gr} | \right) \mathbf{T} | \mathrm{gr} \rangle$, which is induced by the spin-orbit interaction, should be considered.

$$\begin{cases} \delta \langle \mathbf{T} \rangle = -\sum_{\mathrm{ex}} \dfrac{\langle \mathrm{gr} | \mathcal{H}_{\mathrm{SOI}} | \mathrm{exc} \rangle \langle \mathrm{exc} | \mathbf{T} | \mathrm{gr} \rangle}{E_{\mathrm{exc}} - E_{\mathrm{gr}}} + c.c. \\ \phantom{\delta \langle \mathbf{T} \rangle} = \dfrac{2\xi}{\hbar^2} \mathrm{Re} \left[ \sum_{\theta,\sigma_2,\sigma_1} \langle \mu'_1, \sigma_1 | \mathbf{L} \cdot \mathbf{S} | \mu'_2, \sigma_2 \rangle \langle \mu_2, \sigma_2 | \mathbf{T} | \mu_1, \sigma_1 \rangle A(\theta, \sigma_2, \sigma_1) \right] \\ \mathbf{T} = \sum_{\mu_2, \sigma_2; \mu_1, \sigma_1} \mathbf{T}(\mu_2, \sigma_2; \mu_1, \sigma_1) \\ \phantom{\mathbf{T}} = \sum_{\mu_2, \mu_1, \sigma_2, \sigma_1} \langle \mu_2, \sigma_2 | \mathbf{T} | \mu_1, \sigma_1 \rangle \sum_{n_2, n_1, \mathbf{k}} a^*_{n_2, \mu_2, \sigma_1}(\mathbf{k}) a_{n_1, \mu_1, \sigma_1}(\mathbf{k}) c^{\dagger}_{n_2, \sigma_2}(\mathbf{k}) c_{n_1, \sigma_1}(\mathbf{k}) \end{cases} \quad (12)$$

Because $\langle \mathbf{L} \rangle = 0$ for the ground state without spin-orbit interaction, we may set $a_{n,-\mu,\sigma}(\mathbf{k}) = (-1)^{\mu} a^*_{n,\mu,\sigma}(\mathbf{k})$. From this property, the spin-conserved virtual excitation process of the magnetic dipole operator is zero.

$$\left\langle \sum_{\mu_1,\mu_2} \delta T_z(\mu_2, \sigma; \mu_1, \sigma) \right\rangle = 0 \quad (13)$$

Besides this, the spin-flip term is non-zero.

$$\langle \delta T_z(\mu_2,\downarrow;\mu_1,\uparrow)\rangle = 2\operatorname{Re}\left[\sum_{\mu'_1,\mu'_2}\langle \mu'_1,\uparrow|\mathcal{H}_{\text{SOI}}|\mu'_2,\downarrow\rangle\langle \mu_2,\downarrow|\mathbf{T}|\mu_1,\uparrow\rangle A(\theta,\downarrow,\uparrow)\right]$$

$$= 2\operatorname{Re}\left[\sum_{\mu'_1,\mu'_2,\beta}\langle \mu'_1,\uparrow|\mathcal{H}_{\text{SOI}}|\mu'_2,\downarrow\rangle\langle \mu_2,\downarrow|\frac{2}{7\hbar^2}\sum_\beta \frac{L_z L_\beta + L_\beta L_z}{2}S_\beta|\mu_1,\uparrow\rangle A(\theta,\downarrow,\uparrow)\right]$$

$$= \frac{2}{7\hbar}\frac{\mu_2+\mu_1}{2}2\frac{\hbar^2}{\xi}\operatorname{Re}\left[\sum_{\mu'_1,\mu'_2}\langle \mu'_1,\uparrow|\mathcal{H}_{\text{SOI}}|\mu'_2,\downarrow\rangle\langle \mu_2,\downarrow|\mathcal{H}_{\text{SOI}}|\mu_1,\uparrow\rangle A(\theta,\downarrow,\uparrow)\right] \quad (14)$$

From Eq. (14), the following relation can be derived.

$$\delta E_{\text{spin-flip}} = \frac{7}{2}\frac{\xi}{\hbar}\sum_{\mu,\sigma}\frac{1}{2\mu+2\sigma}\langle \delta T_z(\mu+2\sigma,-\sigma;\mu,\sigma)\rangle$$

$$= \frac{7}{2}\frac{\xi}{\hbar}\left(\langle T'_{z,\downarrow\uparrow}\rangle + \langle T'_{z,\uparrow\downarrow}\rangle\right) \quad (15)$$

$$\begin{cases}\langle T'_{z,\downarrow\uparrow}\rangle \equiv \frac{1}{3}\left(\langle \delta T_z(2,\downarrow;1,\uparrow)\rangle - \langle \delta T_z(-1,\downarrow;-2,\uparrow)\rangle\right) + \langle \delta T_z(1,\downarrow;0,\uparrow)\rangle - \langle \delta T_z(0,\downarrow;-1,\uparrow)\rangle \\ \langle T'_{z,\uparrow\downarrow}\rangle \equiv \frac{1}{3}\left(\langle \delta T_z(1,\uparrow;2,\downarrow)\rangle - \langle \delta T_z(-2,\uparrow;-1,\downarrow)\rangle\right) + \langle \delta T_z(0,\uparrow;1,\downarrow)\rangle - \langle \delta T_z(-1,\uparrow;0,\downarrow)\rangle\end{cases}$$

Equation (15) shows the spin-orbit interaction energy from the spin-flip virtual excitation process. It is correlated with the magnetic dipole operator, as suggested by Eq. (28) in Ref. 2. However, Eq. (28) in Ref. 2 and Eq. (15) in this study are not identical, as different approximations are employed in both equations. Specifically, $\langle \mathbf{S}\rangle\cdot\langle \mathbf{T}\rangle \approx \langle \text{gr}|\mathbf{S}\cdot\mathbf{T}|\text{gr}\rangle$ has been employed in Ref. 2 although $\langle \mathbf{S}\rangle\cdot\langle \mathbf{T}\rangle$ and $\langle \text{gr}|\mathbf{S}\cdot\mathbf{T}|\text{gr}\rangle$ are not essentially identical.[25] Note that there is no spin-flip term in $\langle \text{gr}|\mathbf{T}|\text{gr}\rangle$ but it is present in $\langle \text{gr}|\mathbf{S}\cdot\mathbf{T}|\text{gr}\rangle$. Therefore, in this paper, the perturbed ground state was employed to characterize the contributions of the spin-flip terms in $\langle \delta \mathbf{T}\rangle$, as shown in Eq. (12).

Finally, the PMA energy, which is the difference between the spin-orbit interaction energies of perpendicular magnetization and in-plane magnetization, can be expressed as follows from Eqs. (8) and (15).

$$\Delta E \approx -\frac{1}{4}\frac{\xi}{\hbar}\left(\langle \Delta L_{\zeta,\downarrow\downarrow}\rangle - \langle \Delta L_{\zeta,\uparrow\uparrow}\rangle\right) + \frac{7}{2}\frac{\xi}{\hbar}\left(\langle \Delta T'_{\zeta,\downarrow\uparrow}\rangle + \langle \Delta T'_{\zeta,\uparrow\downarrow}\rangle\right) \quad (16)$$

Here, $\langle \Delta L_\zeta\rangle \equiv \langle L_z\rangle - \langle L_x\rangle$ and $\langle \Delta T'_\zeta\rangle \equiv \langle T'_z\rangle - \langle T'_x\rangle$ are used. $\langle L_z\rangle$ and $\langle L_x\rangle$ are evaluated for the

*z*- and *x*- direction components of the spin angular momentum, respectively. The same is the case for $\langle T_z' \rangle$ and $\langle T_x' \rangle$. The orbital magnetic moment ($m_L$)[26] and effective spin magnetic moment ($m_{eff} = m_S - 7m_T$)[27,15] can be characterized by employing the sum-rule analysis[3,26,27] for X-ray magnetic circular dichroism spectroscopy. Usually, the approximation of $\Delta m_{eff} \approx -\Delta 7 m_T$ is employed to estimate the magnetic dipole $T_z$ term ($m_T$). The relation between the orbital magnetic moment anisotropy ($\Delta m_L$) and orbital angular momentum, and the anisotropic part of the magnetic dipole $T_z$ term ($\Delta m_T$) and magnetic dipole operator are expressed as follows:

$$\Delta m_L \equiv m_{L,\perp} - m_{L,//}$$
$$= -\frac{\mu_B}{\hbar}\left(\langle \Delta L_{\varsigma,\uparrow\uparrow}\rangle + \langle \Delta L_{\varsigma,\downarrow\downarrow}\rangle\right), \tag{17}$$

$$\Delta m_{eff} \approx -\Delta 7 m_T$$
$$\equiv -7 m_{T,\perp} - \left(-7 m_{T,//}\right)$$
$$= -7\frac{\mu_B}{\hbar}\left(\langle \Delta T_{\varsigma,\uparrow\uparrow}\rangle + \langle \Delta T_{\varsigma,\downarrow\downarrow}\rangle + \langle \Delta T_{\varsigma,\downarrow\uparrow}\rangle + \langle \Delta T_{\varsigma,\uparrow\downarrow}\rangle\right)$$
$$\approx -\frac{\mu_B}{\hbar}\left(\langle \Delta L_{\varsigma,\uparrow\uparrow}^2\rangle - \langle \Delta L_{\varsigma,\downarrow\downarrow}^2\rangle\right) - \frac{7\mu_B}{\hbar}\left(\langle \Delta T_{\varsigma,\downarrow\uparrow}\rangle + \langle \Delta T_{\varsigma,\uparrow\downarrow}\rangle\right). \tag{18}$$

When strong ferromagnetic materials such as Fe and Co are employed, the orbital angular momentum from the majority spin-band and spin-flip terms can be neglected. Therefore, the PMA energy has the following relation from Eqs. (16) and (17).

$$\Delta E \approx \frac{\lambda}{4\mu_B}\Delta m_L \tag{19}$$

The relation is known as the Bruno model.[1] Equation (19) is reasonable for treating the experimentally obtained PMA energy, for instance, the Au/Co/Au,[15] Fe/MgO,[16] and CoFeB/MgO[17] systems. The VCMA effect in the Fe/Co/MgO system is also explained by Eq. (19).[20] That is, the electric-field-induced change of the orbital magnetic moment anisotropy in the interfacial Co atoms with MgO is quantitatively consistent with the VCMA effect observed in the system.

However, Eq. (19) is not completely applicable when the spin-orbit interaction is large and/or the size of the exchange interaction is small, where significant contributions from the majority spin-band and spin-flip virtual excitation processes would be expected. For instance, in the case of Pt with proximity induced spin-polarization,[10,28-30] $\langle \Delta L_{\varsigma,\uparrow\uparrow} \rangle$, $\langle \Delta T'_{\varsigma,\downarrow\uparrow} \rangle$, and $\langle \Delta T'_{\varsigma,\uparrow\downarrow} \rangle$ cannot be neglected.[10] In Ref. 10, while there is significant contribution of the VCMA effect from the spin-conserved term, there is no significant change in the observed orbital magnetic moment anisotropy in Pt. This strongly suggests that not only $\langle \Delta L_{\varsigma,\downarrow\downarrow} \rangle$ but $\langle \Delta L_{\varsigma,\uparrow\uparrow} \rangle$ has significant contribution to the PMA energy and the measured orbital magnetic moment.

The voltage-induced change in $m_T$ has been observed in Co atoms in the Fe/Co/MgO[20] system and Pt atoms in the Fe/Pt/MgO[10] system. From Eq. (18), the measured $\Delta m_T$ consists of not only $\langle \Delta T_{\varsigma,\downarrow\uparrow} \rangle + \langle \Delta T_{\varsigma,\uparrow\downarrow} \rangle$ but also $\langle \Delta L_{\varsigma,\downarrow\downarrow}^2 \rangle - \langle \Delta L_{\varsigma,\uparrow\uparrow}^2 \rangle$. It should be noted that the spin-conserved terms for $\Delta m_T$ ($\langle \Delta L_{\varsigma,\downarrow\downarrow}^2 \rangle - \langle \Delta L_{\varsigma,\uparrow\uparrow}^2 \rangle$) may possess a finite value even if there are no spin-flip term contributions ($\langle \Delta T_{\varsigma,\downarrow\uparrow} \rangle + \langle \Delta T_{\varsigma,\uparrow\downarrow} \rangle$). In this regard, the observed $\Delta m_T$ of the Co atoms in a strong ferromagnet Fe/Co/MgO system[20] may originate from $\langle \Delta L_{\varsigma,\downarrow\downarrow}^2 \rangle - \langle \Delta L_{\varsigma,\uparrow\uparrow}^2 \rangle$. A finite value of $\Delta m_T$ is evidence of voltage-induced changes in the electric quadrupole, more precisely, in $\langle \mathbf{Q} \cdot \mathbf{S} \rangle$. However, the existence of $\Delta m_T$ is not a sufficient condition for the finite value of the PMA energy induced by the spin-flip terms. In the case of the Fe/Co/MgO system, the influence of $\langle \mathbf{Q} \cdot \mathbf{S} \rangle$ on $\Delta E$ can be neglected because $\langle \Delta T'_{\varsigma,\downarrow\uparrow} \rangle + \langle \Delta T'_{\varsigma,\uparrow\downarrow} \rangle$ is negligibly small.[20]

The situation would be different in the case of Pt atoms in a Fe/Pt/MgO system.[10] As mentioned above, the existence of $\Delta m_T$ shows a voltage-induced change in the electric quadrupole in Pt. The existence ofAs the electric quadrupole strongly suggests the contributions of the spin-flip terms ($\langle \Delta T'_{\varsigma,\downarrow\uparrow} \rangle + \langle \Delta T'_{\varsigma,\uparrow\downarrow} \rangle$) because the exchange split is small in Pt. Moreover, the spin-flip terms from the majority to minority spin bands ($\langle \Delta T'_{\varsigma,\downarrow\uparrow} \rangle$) should be much larger than $\langle \Delta T'_{\varsigma,\uparrow\downarrow} \rangle$. Therefore, $\langle \Delta T'_{\varsigma,\uparrow\downarrow} \rangle$ in Eqs. (16) can be neglected. These discussions are consistent with the first principles study for a Fe/Pt/MgO system (see Fig. 4(c) and (d) in Ref. 10). As mentioned earlier, spin-flip virtual excitation terms are important in strong spin-orbit interaction systems such as Fe/Pt/MgO[10] as well as weak spin-orbit interaction systems such as Fe/MgO[19,21] and Co/Ni.[13] In Refs. 19 and 21, not only the spin-conserved terms but also the spin-flip terms of Fe are important to treat the VCMA effect in Fe/MgO systems. In Ref. 13, the spin-flip terms of Co are one of the dominant sources of PMA energy in Co/Ni systems.

We have constructed an analytic formula to treat the PMA energy in ferromagnetic metals of low symmetry. We found that the anisotropy energy is proportional to a part of the expectation values of the orbital angular momentum and magnetic dipole operator. Specifically, the relation between PMA energy and the measurable physical parameters such as orbital magnetic moment and magnetic dipole $T_z$ term have been revealed via X-ray magnetic circular dichroism spectroscopy. The discussion is consistent with the recent VCMA studies with Co atoms in Fe/Co/MgO and Pt atoms in Fe/Pt/MgO systems.[10,20]

We thank S. Blügel and E. Tamura for the useful discussion. This work was partly supported by the ImPACT program and JSPS KAKENHI (JP18H03880 and JP26103002).